\documentclass[prl,twocolumn,superscriptaddress,nofootinbib]{revtex4-1}

\usepackage{lipsum}
\usepackage{braket}
\usepackage{graphicx}% Include figure files
\usepackage{dcolumn}% Align table columns on decimal point
\usepackage{enumitem}% http://ctan.org/pkg/enumitem, custom enumerate
\usepackage{bm}% bold math
\usepackage{amsmath}
\usepackage{mathtools}
\usepackage{subfigure}
\usepackage{color}
\usepackage{xcolor}
\usepackage{verbatim}
\usepackage{float}

\newcommand{\abs}[1]{\lvert #1 \rvert} % absolute vale
\newcommand{\ev}[1]{\langle #1 \rangle} % expectation value
\newcommand{\lv}[0]{\mathcal{L}} % script L, for Liouville operator
\newcommand{\mD}[0]{\mathcal{D}} % script D, for Lindblad terms

\newcommand{\OQ}[0]{\Omega_{\text{QND}}} % omega_QND
\newcommand{\vis}[0]{\mathcal{V}} % ''visibility''
\newcommand{\phiApp}[0]{\ensuremath{\phi^{\left(\text{a}\right)}}} % script D, for Lindblad terms

\begin{document}

\title{Continuous real-time tracking of a quantum phase below the standard quantum limit}
\author{Athreya Shankar}
\email{athreya.shankar@colorado.edu}
\author{Graham P. Greve}
\author{Baochen Wu}
\author{James K. Thompson}
\author{Murray Holland}
\affiliation{JILA, NIST, and Department of Physics, University of Colorado, 440 UCB, 
Boulder, CO  80309, USA}
\date{\today}

\begin{abstract}
    We propose a scheme for continuously measuring the evolving quantum phase of a collective spin composed of $N$ pseudospins. Quantum non-demolition measurements of a lossy cavity mode interacting with an atomic ensemble are used to directly probe the phase of the collective atomic spin without converting it into a population difference. Unlike traditional Ramsey measurement sequences, our scheme allows for real-time tracking of time-varying signals. As a bonus, spin-squeezed states develop naturally, providing real-time phase estimation significantly more precise than the standard quantum limit of $\Delta \phi_\text{SQL} = 1/\sqrt{N}$ radians.
\end{abstract}

\maketitle

Quantum systems have become robust platforms for metrology and tests of fundamental physics. Many applications rely on the dynamics of pseudospin-1/2 systems with two long-lived quantum states, $\ket{\uparrow}$ and $\ket{\downarrow}$. After preparing an equal superposition of these two states, a physical interaction is studied by investigating its effect on the relative phase $\phi(t)$, with the state of each spin evolving in time as $\ket{\psi(t)}=\left(\ket{\downarrow}+e^{i\phi(t)}\ket{\uparrow}\right)/\sqrt{2}$. We propose a novel scheme that enables continuous tracking of this relative phase. Our scheme continuously and directly measures the real-time phase $\phi(t)$ unlike the widely used Ramsey sequence \cite{ramsey1950PR,sr_clock, yb_clock, katori_clocks,bouyer_review, tino_g, rasel_freefall,polzik_magnetometry,reidel_darkmatter, muller_darkenergy, new_grav_wave_method,garttner2017Nat}, which indirectly measures the net accumulated phase $\phi(T)$ during an interrogation time $T$. The typically destructive readout in a Ramsey sequence requires multiple state resets, rotations and repetitions of the sequence to infer the phase at different times from a population difference. In contrast, a single run of our protocol yields a continuous time series of phase measurements. Therefore, our scheme enables real-time tracking of time-varying signals that are not reproducible.

As an added benefit, our scheme yields continuous phase estimates with precision well beyond the standard quantum limit (SQL) of $\Delta \phi_\text{SQL} = 1/\sqrt{N}$ radians that limits readout precision with $N$ unentangled spins. In comparison to several proposals and experiments \cite{kitagawa1993PRA,leroux2010PRL,schleierSmith2010PRA,kuzmich2000PRL,bohnet2014Nat,cox2016PRL,hosten2016Nat} that have demonstrated squeezed states with precision beyond the SQL, our scheme enjoys the advantage that the squeezing is produced, the phase accumulated, and the readout performed, all in the same spin quadrature. 

Recent experiments have demonstrated phase tracking of a spin using quantum non-demolition (QND) measurements via a Faraday rotation angle \cite{colangelo2017Nat}. In contrast, our proposal is based on interfering Raman transitions in a cavity and enables an intuitive interpretation of phase tracking in terms of elementary atom-cavity interactions that nearly balance one another. 
Our scheme directly reveals a phasor precessing in the equatorial plane of a Bloch sphere, in the spirit of the ``hand on a clock'' analogy at the core of quantum metrology.

\begin{figure}[!htb]
    \centering
    \includegraphics[width=\columnwidth]{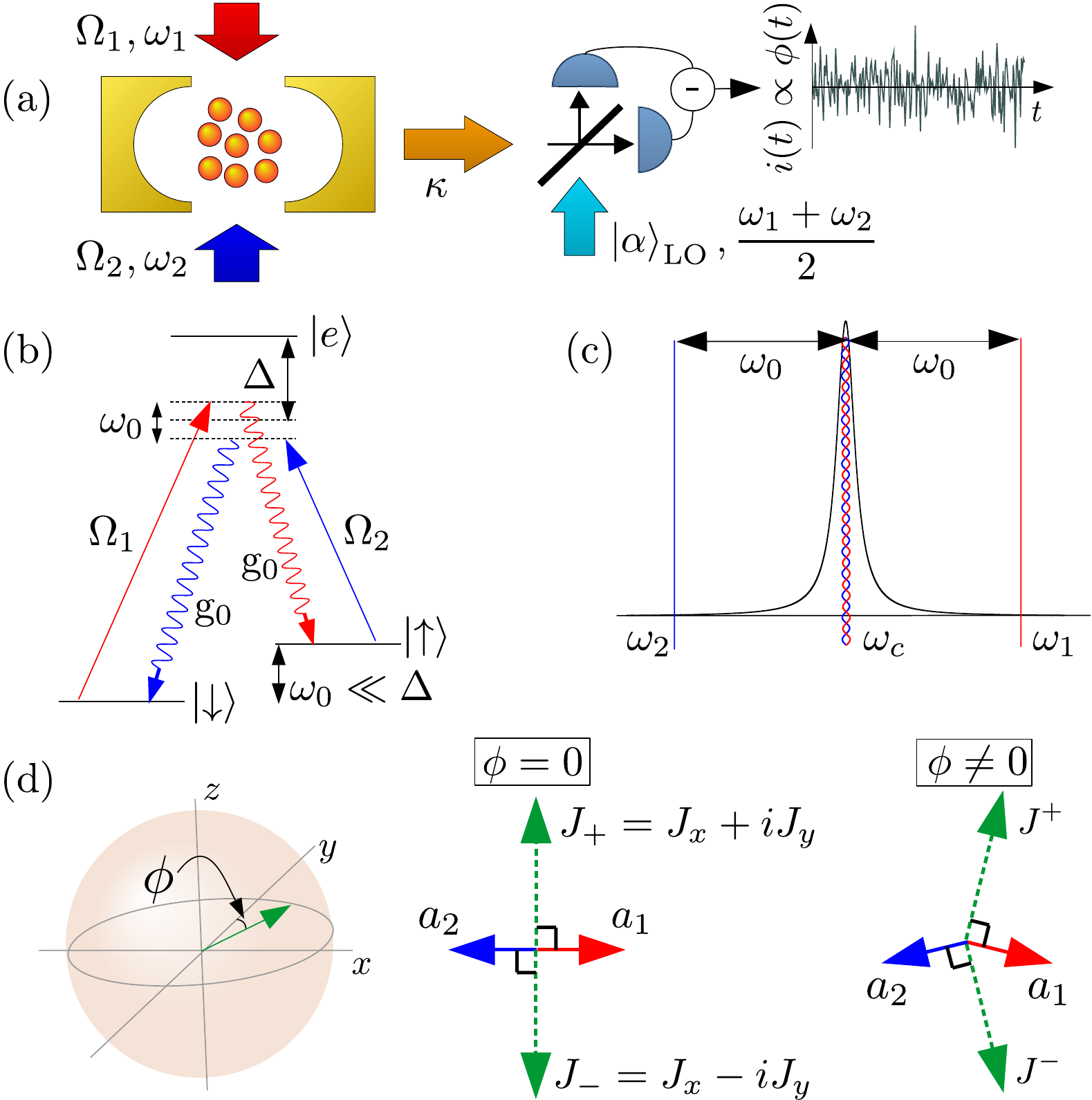}
    \caption{ Schematic and working principle. (a) Two lasers drive a collection of atoms to interact with a cavity mode. The relative phase $\phi(t)$ can be continuously tracked by homodyne detection of the field leaking out. (b) Cavity-assisted Raman transitions: The red (blue) pathway leads to the emission of a cavity photon accompanied by a spin flip $\ket{\downarrow}\rightarrow\ket{\uparrow}$ ($\ket{\uparrow}\rightarrow\ket{\downarrow}$). (c) Hierarchy of frequencies. (d) Classical Bloch vector picture: The red and blue pathways set up balanced, opposing superradiance pathways that lead to a coherent cancellation of the intracavity field when the Bloch vector (green) is along the $y$-axis ($\phi=0$). When the Bloch vector has a small $x$-component ($\phi \neq 0$), the intracavity field from the two pathways add constructively, giving rise to non-zero output field.} 
    \label{fig:expt_setup}
\end{figure}

We represent the collective angular momentum of $N$ atomic spins by a classical Bloch vector of length $N/2$ with components $J_x$, $J_y$, $J_z$ (Fig. \ref{fig:expt_setup}(d, left)). With all spins initially in the same equal superposition state, the Bloch vector lies in the equatorial plane along a direction that we define as the $y$-axis. As the phase evolves, the Bloch vector acquires a small $x$-component, $J_x = \frac{N}{2} \sin \phi(t) \approx \frac{N}{2} \phi(t)$, for small deflections, and we propose a straightforward extension to large deflections in the conclusion. We  arrange atom-cavity interactions wherein a cavity field quadrature is sourced by $J_x$. Continuous homodyne detection of this quadrature amounts to real-time, continuous, QND measurement of $\phi(t)$. 

We consider $N$ atoms trapped at the antinodes of a cavity with resonance frequency $\omega_c$ and decay rate $\kappa$, as shown in Fig.~\ref{fig:expt_setup}(a). The states $\ket{\downarrow}$ and $\ket{\uparrow}$ have an energy separation $\hbar \omega_0 \gg \hbar \kappa$ and form a pseudospin-1/2 system described by the Pauli spin operators $\hat{\sigma}_i, \; i=x,y,z$, with raising (lowering) operators $\hat{\sigma}_+$ ($\hat{\sigma}_-$). The $N$ atoms form a collective spin with total angular momentum components $\hat{J}_x,\hat{J}_y,\hat{J}_z$, with $\hat{J}_i = \sum_{j=1}^N \hat{\sigma}_i^j/2$. We assume the dipole-allowed transitions $\ket{\downarrow}\leftrightarrow\ket{e}$ and $\ket{\uparrow}\leftrightarrow\ket{e}$ with frequencies $\omega_{\downarrow e}$ and $\omega_{\uparrow e}$ to be respectively driven using lasers with frequencies $\omega_1$ and $\omega_2$ in a far-detuned regime with detuning $\Delta \gg \omega_0,\kappa$, allowing for the adiabatic elimination of $\ket{e}$ \cite{james2007CJP}. The two drive lasers differ by a frequency $2 \omega_0$ (Fig.~\ref{fig:expt_setup}(c)) and do not by themselves drive $\ket{\downarrow} \leftrightarrow \ket{\uparrow}$ Raman transitions; however, they are symmetrically detuned by $\omega_0$ from $\omega_c$ and participate in cavity-assisted Raman transitions as illustrated in Fig.~\ref{fig:expt_setup}(b) \cite{suppMat}. When the Rabi frequencies of the two drive lasers are balanced, i.e. $\Omega_1=\Omega_2=\Omega_0$, the atom-cavity Hamiltonian, to leading order in $1/\Delta$, is simply the sum of a Jaynes-Cummings and an anti-Jaynes-Cummings interaction and is given by \cite{suppMat}
\begin{equation}
    \hat{H}_{\text{QND}} = \frac{\hbar\Omega_{\text{QND}}}{2} \hat{X}\hat{J}_x.
    \label{eqn:qnd_hamiltonian}
\end{equation}
Here $\hat{X}=(\hat{a}+\hat{a}^\dagger)/\sqrt{2}$ is the amplitude quadrature, with $\hat{a},\hat{a}^\dagger$ the annihilation and creation operators for the cavity mode, and $\hat{Y} = (\hat{a}-\hat{a}^\dagger)/\sqrt{2}i$ is the conjugate phase quadrature such that $[\hat{X},\hat{Y}]=i$. The atom-cavity interaction strength is $\Omega_{\text{QND}} = \sqrt{2}\Omega_0 g_0/\Delta$ with $g_0$ the single atom-cavity vacuum Rabi frequency. If the two drive lasers have initial phases $\psi_1$ and $\psi_2$, the cavity quadrature $\left(\hat{a}^\dag e^{i(\psi_1+\psi_2)/2}+\;\text{H.c.}\right)$ is coupled to the spin component $\left(\hat{J}_+ e^{i(\psi_1-\psi_2)/2}+\;\text{H.c.}\right)$, where $\hat{J}_+=\hat{J}_x+i\hat{J}_y$. Here we assume $\psi_1=\psi_2=0$ without loss of generality. 

Classically, the intracavity fields established by the two balanced drives exactly cancel when $J_x = 0$ (Fig. \ref{fig:expt_setup}(d)). However, even with $\ev{\hat{J}_x}=0$, $\ev{\hat{J}_x^2}\neq 0$, i.e. quantum fluctuations source the $Y$ quadrature of the cavity field. In the regime $\kappa^2 \gg N\Omega_{\text{QND}}^2$, $\hat{Y}$ is slaved to $\hat{J}_x$ as 
\begin{equation}
    \hat{Y}(t) \approx -\frac{\Omega_{\text{QND}}}{\kappa}\hat{J}_x (t) + \hat{\mathcal{F}}(t),
    \label{eqn:cavity_slaving}
\end{equation}
\noindent where the noise operator $\hat{\mathcal{F}}(t)$ arises from coupling of the cavity mode to external modes through the lossy mirror (Fig.~\ref{fig:expt_setup}(a)) \cite{xu2016PRL,meystre1998elements}. The field leaking out is to be monitored via balanced homodyne detection using a local oscillator at frequency $(\omega_1 + \omega_2)/2$ with phase tuned to detect the output field quadrature that is sourced by the intracavity $Y$ quadrature. The photocurrent thus recorded is a measurement of the $Y$ quadrature which, from Eq.~(\ref{eqn:cavity_slaving}), amounts to measuring $J_x$.

Measurement back-action in the $J_z$ quadrature arises because of the indistinguishability of the two pathways that give rise to the intracavity field (Fig. \ref{fig:expt_setup}(b)): The field leaking out is consistent with equal probability amplitudes for tipping the Bloch vector above or below the equator and therefore increases the spread in $J_z$ without affecting its mean value.

The drive lasers also lead to undesirable, off-resonant free-space scattering processes with total rate $\gamma_\text{sc}$ that degrade atomic coherence. We consider three such single-atom decoherence mechanisms \cite{suppMat}: (a) dephasing with probability $r_d$: random rotation about the $z$-axis, (b) spontaneous Raman spin flips : $\ket{\downarrow} \rightarrow \ket{\uparrow}$ ($r_{\downarrow \uparrow}$) and $\ket{\uparrow} \rightarrow \ket{\downarrow}$ ($r_{\uparrow \downarrow}$), and (c) atom loss ($r_l$): the atom decays to a state $\ket{s}$ outside the $\ket{\downarrow}-\ket{\uparrow}$ manifold and no longer interacts with the cavity mode. The probabilities are related by $r_d + r_{\downarrow \uparrow} + r_{\uparrow \downarrow} + r_l = 1$.

Under continuous measurement, the dynamics of the density matrix $\rho$ of the atom-cavity system is governed by the stochastic master equation \cite{wiseman1993PRA, bowen2015book,wiseman2010quantum}:
\begin{eqnarray}
\dot{\rho} = & -i/\hbar[\hat{H}_\text{QND},\rho] + \kappa \mD[\hat{a}]\rho + \gamma_\text{sc} \sum_{j=1}^N \lv_1^j\rho \nonumber\\
             & + \sqrt{\eta \kappa} \xi(t) \left(i\rho\hat{a}^\dag -i\hat{a}\rho -\sqrt{2}\ev{\hat{Y}}\rho \right),
             \label{eqn:stoch_master_eqn}
\end{eqnarray}

\noindent with decoherence effects bundled in $\lv_1^j \rho$, given by 
\begin{eqnarray}
    \lv_1^j \rho & = & r_{\downarrow \uparrow} \mD[\hat{\sigma}_+^j]\rho
               + r_{\uparrow \downarrow} \mD[\hat{\sigma}_-^j]\rho 
                + \frac{r_d}{4} \mD[\hat{\sigma}_z^j]\rho \nonumber\\
               & + &\frac{r_l}{2} \left( \mD\left[\ket{s}_j\bra{\downarrow}_j\right] \rho + \mD\left[\ket{s}_j\bra{\uparrow}_j\right] \rho \right),  
    \label{eqn:fss_terms}
\end{eqnarray}

\noindent with $\mD[\hat{O}]\rho = \hat{O}\rho\hat{O}^\dag - \hat{O}^\dag \hat{O}\rho/2 - \rho \hat{O}^\dag \hat{O}/2$, the Lindblad dissipator. In Eq.~(\ref{eqn:stoch_master_eqn}), $\eta$ is the detection efficiency, and $\xi(t)$ is a white-noise process satisfying $\overline{\xi(t)}=0$ and $\overline{\xi(t)\xi(t')}=\delta(t-t')$. The measured photocurrent $i(t)$ is  
\begin{equation}
    i(t) =  \mathcal{G} e \abs{\alpha_\text{LO}} \left( \eta\sqrt{2\kappa}\ev{\hat{Y}} + \sqrt{\eta}\xi(t) \right),
    \label{eqn:photocurrent}
\end{equation}

\noindent with detector gain $\mathcal{G}$, electronic charge $e$, and local oscillator photon flux $\abs{\alpha_\text{LO}}^2$ with units of photons/time.

With no decoherence, measuring for very long times will result in preparing states arbitrarily close to Dicke states in the $J_x$ basis. However, decoherence restricts the maximum achievable squeezing well before the state begins to wrap around the Bloch sphere. This enables a Gaussian approximation where we only track the dynamics of the means and covariances of all operators and pairs of operators of the atom-cavity system. The 5 operators $\hat{X},\hat{Y},\hat{J}_x,\hat{J}_y$ and $\hat{J}_z$ result in a total of 20 dynamical equations \cite{suppMat}.

We average the simulated photocurrent (Eq.~(\ref{eqn:photocurrent})) in a window $[T_\text{i},T_\text{f}]$ to obtain an estimate as
\begin{equation}
    J_x^\text{(m)} = -\frac{\kappa}{\Omega_\text{QND}}Y^\text{(m)} = \frac{-(\mathcal{G} e\abs{\alpha_\text{LO}})^{-1}}{\eta \sqrt{C\gamma_\text{sc}}(T_\text{f} - T_\text{i})} \int_{T_\text{i}}^{T_\text{f}}  i(t) dt,
    \label{eqn:jx_estimate}
\end{equation}

\noindent where $C = 2\Omega_\text{QND}^2/\kappa \gamma_\text{sc}$ is the dimensionless atom-cavity cooperativity \cite{suppMat}. The phase is estimated as $\phi^\text{(m)}=(J_x^\text{(m)}/(N/2))/\mathcal{V}(t)$, where the visibility $\mathcal{V}(t)$ \cite{visibility} accounts for the shortening of the Bloch vector, evaluated either at the window center or end depending on where the phase is estimated. While we use a simple averaging procedure for clarity, optimal filters, such as Kalman filters, can be applied for superior phase tracking \cite{geremia2003PRL,stockton2004PRA,bouten2007SIAM}.

The precision of the phase estimate in a window is determined by the window duration. A characteristic time, $T_0 = (\eta C \gamma_\text{sc})^{-1}/(N/4)$, is the time required to average down the photon shot-noise ($\xi(t)$ term, Eq.~(\ref{eqn:photocurrent})) in estimating $J_x^\text{(m)}$ (Eq.~(\ref{eqn:jx_estimate})) to the standard quantum limit $\Delta J_{x,\text{SQL}}^2=N/4$.

\begin{figure}[ht]
    \centering
    \includegraphics[width=\columnwidth]{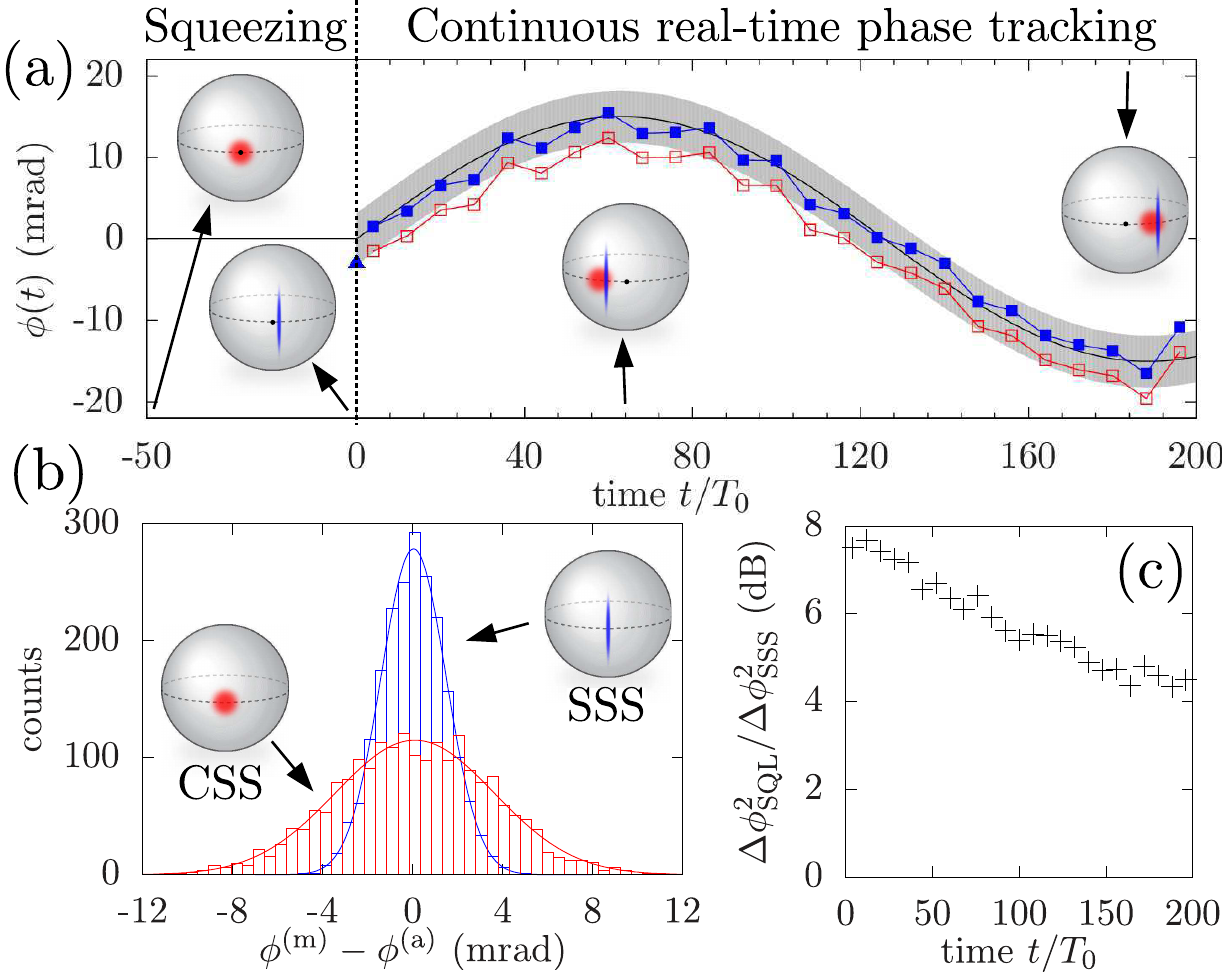}
    \caption{Real-time continuous tracking of a time-varying phase. (a) A single experimental run: A squeezed state is prepared during $[-50T_0,0]$, with the initial measured phase $\phi^\text{(m)}_0$ (blue triangle) varying in each run. Subsequently, a phase modulation $ \phiApp(t) = 15 \; \text{mrad} \times \sin (t/40T_0)$ (black line) is applied e.g. using a time-varying magnetic field. The blue, filled (red, hollow) markers are estimates  $\phi^\text{(m)}_\text{SSS}$ ($\phi^\text{(m)}_\text{CSS}$) of the phase using the measured photocurrent in windows of duration $8 T_0$ that account for (do not account for) the initial offset $\phi^\text{(m)}_0$. The gray shaded region indicates the $1\text{-}\sigma$ SQL tolerance for this applied signal. Representative Bloch spheres for $t \le 0$ indicate the state before and after the state preparation stage. For $t>0$, Bloch spheres indicate the deflection of the spin as a result of the phase modulation (black dots on the spheres indicate the zero phase reference), as well as the equivalent spin state used for the respective estimates $\phi^\text{(m)}_\text{CSS}$, $\phi^\text{(m)}_\text{SSS}$. (b) Histogram of phase errors $\phi^\text{(m)}_\text{SSS}-\phiApp$ (blue) and $\phi^\text{(m)}_\text{CSS}-\phiApp$ (red) over 2048 runs in one particular measurement window [$48 T_0, 56 T_0$]. (c) Single-run precision gain of the estimates $\phi^\text{(m)}_\text{SSS}$ relative to the SQL at different window centers $t$. Here, $\Delta \phi^2_\text{SSS}$ is the variance of Gaussian fits to histograms such as the blue histogram in (b). Decoherence results in decreased gain over time.}
    \label{fig:real_time_tracking}
\end{figure}

For our numerical experiments, we use $N = 10^5$ atoms identically coupled to a cavity mode with $C=0.1$. We work in a bad-cavity regime such that $NC\gamma_\text{sc} = 0.2\kappa$, achievable by arranging for $\Omega_\text{QND} = 10^{-3}\kappa$. We adopt a ``symmetric loss'' model wherein the three decoherence mechanisms degrade the atomic coherence at equal rates, and spin-flips in either direction occur with equal probability. This implies $r_d = 1/3$, $r_{\downarrow \uparrow}=r_{\uparrow \downarrow}=1/6$ and $r_l=1/3$. Our results are not very sensitive to the specific choice of relative rates. The loss in visibility only depends on the total decoherence rate, while the measurement of $J_x$ marginally improves if the atom loss channel is dominant (see Eq.~(\ref{eqn:analytic_var})). Finally, the detection efficiency is assumed to be $\eta=0.4$ \cite{cox2016PRL}. 

We now demonstrate the ability of our scheme to track in real-time, a phase modulation $\phiApp(t)$ applied for $t>0$ (Fig.~\ref{fig:real_time_tracking}(a, black solid line)). At time $t = -50T_0$, the collective spin is initialized to a coherent spin state (CSS) along the $y$-axis whose initial phase is $\phi_0 = 0$.
 First, measuring the photocurrent in the state preparation window $[-50T_0,0]$ gives a phase estimate $\phi^\text{(m)}_0$ (blue triangle). This estimate is obtained at the end of this window using the procedure described below Eq.~(\ref{eqn:jx_estimate}). The value of $\phi^\text{(m)}_0$ varies from trial-to-trial with a variance $\Delta \phi^2_\text{SQL} = 1/N$ corresponding to the phase uncertainty of the initial CSS. The long state preparation window ensures strong averaging down of the photon shot-noise, leading to a state with reduced phase uncertainty around $\phi^\text{(m)}_0$, i.e. a spin squeezed state (SSS). For the subsequent real-time tracking, two choices for the initial phase reference could be used: $\phi_0(=0)$ or $\phi^\text{(m)}_0$.

%The collective spin is initialized to a coherent spin state (CSS) along the $y$-axis at time $t = -50T_0$ whose initial phase is $\phi_0 = 0 \pm \Delta\phi_\text{SQL}$.

During the time $[0,200T_0]$, we average the photocurrent in windows of duration $8T_0$ to extract a raw phase estimate $\phi^\text{(m)}(j)$ for window $j=1, 2, \ldots$. We construct two estimates for the phase at the window centers, $\phi^\text{(m)}_\text{CSS}(j)=\phi^\text{(m)}(j)-\phi_0$ (hollow red squares), and $\phi^\text{(m)}_\text{SSS}(j)=\phi^\text{(m)}(j)-\phi^\text{(m)}_0$ (filled blue squares). The precision of these estimates is determined not just by the window duration over which the raw estimate is obtained, but also by the precision of the phase reference. To determine the single-run precision of these estimates, we run 2048 trials of the experiment and histogram the error in these estimates, an example of which is shown in Fig.~\ref{fig:real_time_tracking}(b) for the window [$48 T_0, 56 T_0$]. The estimates $\phi^\text{(m)}_\text{CSS}$ use the imprecise zero phase $\phi_0$ of the initial CSS as reference, and lead to a broad error histogram (red). In contrast, the estimates $\phi^\text{(m)}_\text{SSS}$ lead to a narrow error histogram (blue) whose spread is instead dominated by the imprecision in obtaining the raw estimates $\phi^\text{(m)}(j)$ over short windows (here, $8T_0$), demonstrating the improved precision of the phase reference $\phi^\text{(m)}_0$ over $\phi_0$ \cite{dcErrorNote}. In Fig.~\ref{fig:real_time_tracking}(c) we show that the variance $\Delta \phi_\text{SSS}^2$ of the estimates $\phi^\text{(m)}_\text{SSS}$ is significantly less than $\Delta \phi_\text{SQL}^2$ in all windows over the time we consider here, demonstrating the potential for real-time phase tracking with precision beyond the SQL.

\begin{figure}[htb]
    \centering
    \includegraphics[width=\columnwidth]{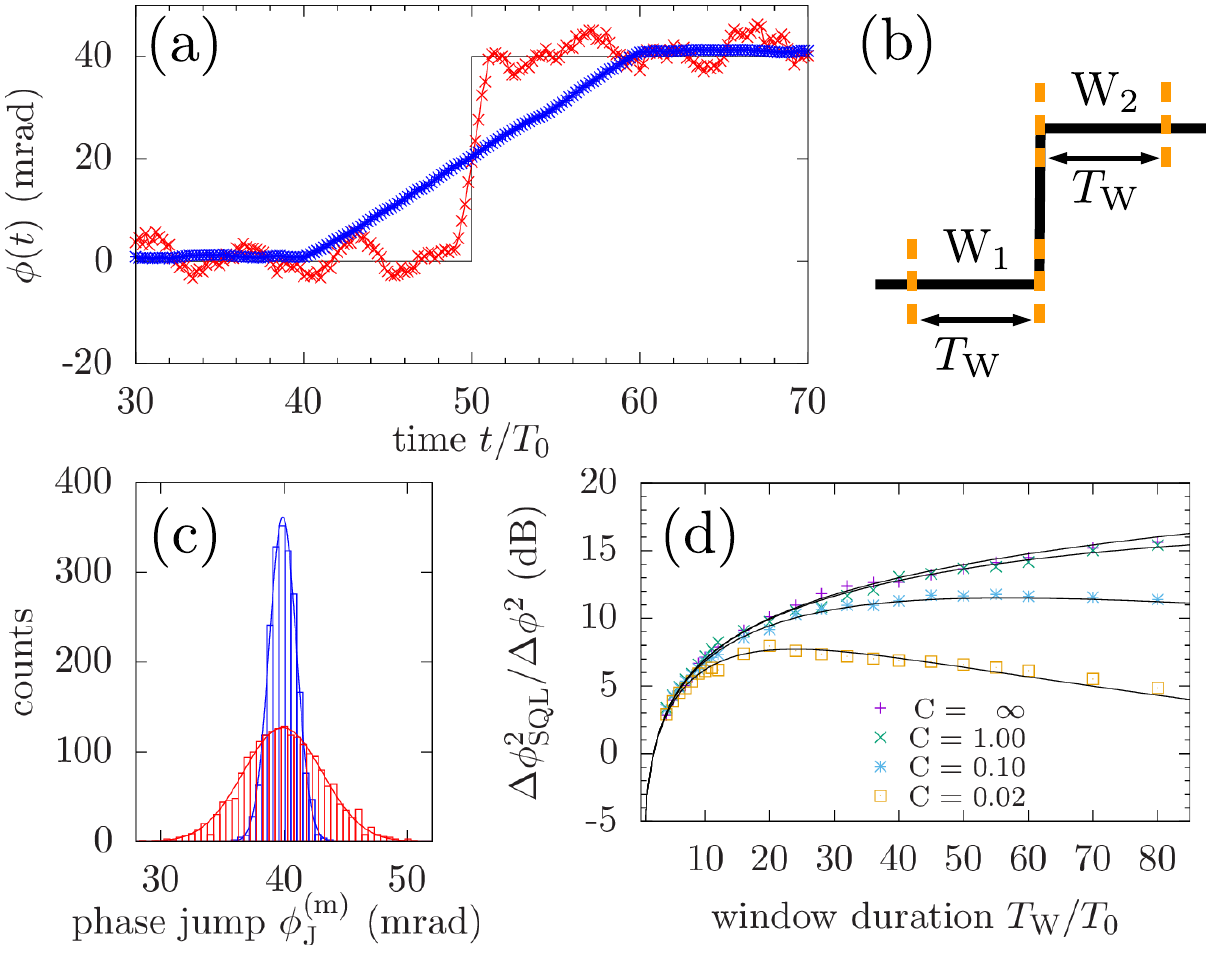}
    \caption{(a) A sudden jump in the phase with amplitude $\phi_\text{J} = 40 \; \text{mrad}$ at $T_\text{J} = 50 T_0$ is tracked in the same run using moving windows of durations $T_\text{W} = 2T_0$ (red) and $T_\text{W}=20T_0$ (blue), showing the faster response of the shorter window. (b) Protocol to estimate $\phi_\text{J}$. (c) Histograms, over 2048 runs, of $\phi^\text{(m)}_\text{J}$ for $T_\text{W} = 2T_0$ (red) and $T_\text{W} = 20T_0$ (blue), demonstrating the greater precision of the longer window. For $T_\text{W} = 2T_0$, $\text{W}_2$ was offset by a small time $0.2T_0$ to allow transients on timescales of $\kappa^{-1}$ to decay. (d) Gain in precision over a CSS in Ramsey mode as the duration of $\text{W}_1$ and $\text{W}_2$ is varied, for fixed $C\gamma_\text{sc}$ and different values of $C$. Analytic results (lines) calculated using Eqs.~(\ref{eqn:metrological_gain}) and~(\ref{eqn:analytic_var}) are in excellent agreement with simulations (markers).}
    \label{fig:sudden_jump_ramsey}
\end{figure}

An advantage of our scheme is that the same photocurrent data from a single run can be analyzed using multiple methods to extract varying information. As a demonstration, we use varying window durations $T_\text{W}$ to extract precise timing and amplitude information from a sudden jump in phase (at $T_\text{J}=50T_0$ in Fig.~\ref{fig:sudden_jump_ramsey}(a)). Starting with an initial CSS at $t=0$, we continuously estimate the phase by averaging the photocurrent over moving windows of durations $T_\text{W} = 2T_0$ (red) and $T_\text{W} = 20T_0$ (blue). Clearly, the shorter window reproduces the time variation of the phase more precisely.  To estimate the amplitude of the jump $\phi_\text{J}$, we compute the difference $\phi^\text{(m)}_\text{J}$ in the estimates $\phi^\text{(m)}_{\text{W}_1}, \phi^\text{(m)}_{\text{W}_2}$ in the two windows $\text{W}_1\equiv[T_\text{J}-T_\text{W},T_\text{J}]$ and $\text{W}_2\equiv[T_\text{J},T_\text{J}+T_\text{W}]$ that border the jump time $T_\text{J}$ (Fig.~\ref{fig:sudden_jump_ramsey}(b)) \cite{jumpTimeNote}. While the shorter window results in faster response, the longer window gives a more precise estimate of the jump amplitude (Fig.~\ref{fig:sudden_jump_ramsey}(c)). 

Alternatively, the sudden phase jump in the protocol depicted in Fig.~\ref{fig:sudden_jump_ramsey}(b) can be replaced with a ``dark'' phase accumulation time of duration $T_\text{D}$ where no measurements are performed. The scheme can then be identified as a Ramsey-like sequence where a squeezed state is prepared in $\text{W}_1$, phase accumulates in an interrogation time $T_\text{D}$, and finally, phase is read out in $\text{W}_2$, without ever converting the phase information into a population difference. In this Ramsey mode, the achievable gain in phase resolution using the prepared squeezed state compared to a CSS is
\begin{equation}
    \frac{\Delta \phi^2_\text{SQL}}{\Delta \phi^2} = \frac{\Delta J_{x,\text{SQL}}^2}{\left(\Delta J_{x,\text{diff.}}^\text{(m)}\right)^2} \mathcal{V}^2,
    \label{eqn:metrological_gain}
\end{equation}

\noindent where $J_{x,\text{diff}}^\text{(m)} = J_{x,W_2}^\text{(m)} - J_{x,W_1}^\text{(m)}$ and $\mathcal{V}$ is the visibility at the end of the first window \cite{bohnet2014Nat, cox2016PRL}. Fig.~\ref{fig:sudden_jump_ramsey}(d) plots the numerically extracted gain (markers) versus the window duration $T_\text{W}$ for different values of cooperativity $C$. Gaussian fits to histograms of $J_{x,\text{diff.}}^\text{(m)}$ were used to extract values for $(\Delta J_{x,\text{diff.}}^\text{(m)})^2$. We find analytically that the normalized variance in the difference measurement varies with $T_\text{W}$ as \cite{suppMat}
\begin{equation}
    \frac{(\Delta J_{x,\text{diff.}}^\text{(m)})^2(T_\text{W})}{\Delta J_{x,\text{SQL}}^2} = 2\frac{T_0}{T_\text{W}} + \frac{8\beta}{3\eta NC}\frac{T_\text{W}}{T_0}, 
    \label{eqn:analytic_var}
\end{equation}

\noindent where $\beta = r_d + r_{\downarrow \uparrow} + r_{\uparrow \downarrow} + r_l/2$, giving a minimum normalized variance of $8\sqrt{\beta/3\eta NC}$ at $T_{\text{W}}^\text{opt}=T_0 \sqrt{3\eta N C/4\beta}$. The expression for $\beta$ shows that the normalized variance is not very sensitive to the relative probabilities of the decoherence mechanisms. For typical values of $C\sim 0.1$ and $N\sim 10^5$, Fig.~\ref{fig:sudden_jump_ramsey}(d) shows that  a gain upwards of $11 \; \text{dB}$ can be achieved. The $(NC)^{-1/2}$ scaling of the minimum normalized variance in $J_{x,\text{diff}}^\text{(m)}$ leads to an optimal phase resolution scaling as $\Delta\phi \sim N^{-3/4}$ compared to $\Delta \phi_\text{SQL} = N^{-1/2}$ radians.

In conclusion, we have proposed and analyzed a scheme for continuous real-time tracking of a quantum phase with precision beyond the SQL. Interfering cavity-assisted Raman transitions have been considered previously for deterministic squeezing schemes \cite{sorensen2002PRA} and quantum simulations of the Dicke model \cite{dimer2007PRA,baumann2010Nat,zhang2018PRA}. The frequency arrangement of our drive lasers is also related to two-tone drive schemes for back-action evading measurements of mechanical oscillators \cite{braginsky1980Science,clerk2008NJP,bowen2015book} and for measuring the state of individual superconducting qubits \cite{hacohen-Gourgy2016Nat,eddins2018PRL}. Furthermore, while Ramsey sequences only measure phase changes unambiguously in the interval $[-\pi/2,\pi/2]$, our scheme readily extends to tracking large excursions $\abs{\phi(t)} \gg \pi$: The measured current $i(t)$ can be used in a feedback loop \cite{wiseman1995PRL,armen2002PRL,wiseman1993PRL,wiseman1994PRA,thomsen2002JPhysB} to adjust the differential phase offset $\psi_1-\psi_2$ of the drive lasers such that $i(t)$ is continuously driven back to zero. The feedback loop continuously adjusts the spin component probed by the cavity mode such that it is always perpendicular to the mean spin direction, while mapping the phase $\phi(t)$ onto the feedback signal as $\phi(t) = (\psi_1-\psi_2)/2$. This way, large phase excursions can be tracked while remaining in the small angle measurement limit, also greatly suppressing sensitivity to variations or uncertainties in scale factors relating $i(t)$ to $\phi(t)$, including uncertainties in atom number \cite{suppMat}. By encoding the spins in hyperfine levels that have an intrinsic splitting, our scheme has the unique capability to greatly increase the unambiguous interval of phase evolution that can be continuously tracked, for example in atomic clocks. While feedback schemes using intermittent non-demolition population measurements have been used to extend this interval in a Ramsey-like sequence \cite{kohlhaas2015PRX}, our scheme continuously tracks the phase and removes the need for state rotations altogether. It will be interesting to see if this scheme can be adapted to optical clock transitions, perhaps in ${}^{87}$Sr.

\begin{acknowledgments}
We would like to thank Chengyi Luo, John Cooper, Nicola Poli, John Teufel and Graeme Smith for fruitful discussions. This work was supported by NSF PFC grant number PHY 1734006, DARPA Extreme Sensing, and NIST.
\end{acknowledgments}

\bibliographystyle{apsrev4-1}

%\clearpage
%\appendix

%%%%%%%%%%%%%%%%%%%%%%%%%%%%%%%%%%%%%%%%%%%%%%%%%%%%%%%%%%%%%%%%%%%%%%%%%%%%%%%%%%%%%%%%%%%%%%
%SUPPLEMENTAL MATERIAL

\clearpage

\begin{comment}
\onecolumngrid
\begin{center}
  \textbf{\large Supplemental Material: Continuous real-time tracking of a quantum phase below the standard quantum limit}\\[.2cm]
  Athreya Shankar$^{1,*}$, Graham P. Greve$^1$, Baochen Wu$^1$, James K. Thompson$^1$ and Murray Holland$^1$\\[.1cm]
  {\itshape ${}^1$JILA, NIST, and Department of Physics, University of Colorado, 440 UCB, 
Boulder, CO  80309, USA\\}
${}^*$Electronic address: athreya.shankar@colorado.edu\\
(Dated: \today)\\[1cm]
\end{center}
\twocolumngrid
\end{comment}

\onecolumngrid
\begin{center}
  \textbf{\large Supplemental Material: Continuous real-time tracking of a quantum phase below the standard quantum limit}\\[.2cm]
  Athreya Shankar$^{1,*}$, Graham P. Greve$^1$, Baochen Wu$^1$, James K. Thompson$^1$ and Murray Holland$^1$\\[.1cm]
  {\itshape ${}^1$JILA, NIST, and Department of Physics, University of Colorado, 440 UCB, 
Boulder, CO  80309, USA\\}
\end{center}
\twocolumngrid

\setcounter{equation}{0}
\setcounter{figure}{0}
\setcounter{table}{0}
\setcounter{page}{1}
\renewcommand{\theequation}{S\arabic{equation}}
\renewcommand{\thefigure}{S\arabic{figure}}
\renewcommand{\bibnumfmt}[1]{[S#1]}
\renewcommand{\citenumfont}[1]{S#1}

We refer to the equations in the Main Text using regular arabic numerals and prefix the equations introduced in the Supplemental Material with the letter `S'.

\section{Adiabatic elimination of the excited state}

The Hamiltonian for the interaction of the atoms with the drive lasers and cavity mode is ($\hbar=1$)
\begin{eqnarray}
&&\hat{H} = \omega_c \hat{a}^\dag \hat{a} 
          - \omega_{\downarrow e} \sum_j \ket{\downarrow}_j\bra{\downarrow}
          - \omega_{\uparrow e} \sum_j \ket{\uparrow}_j\bra{\uparrow} \nonumber\\
&&\hphantom{\hat{H} } + \sum_j \left( \frac{\Omega_1}{2}\ket{e}_j\bra{\downarrow}e^{-i\omega_1 t}                    
+\frac{\Omega_2}{2}\ket{e}_j\bra{\uparrow}e^{-i\omega_2 t} + \; \text{H.c.}\right) \nonumber\\
&&\hphantom{\hat{H} } + \sum_j \left( \frac{g_1}{2}\hat{a}\ket{e}_j\bra{\downarrow} 
+ \frac{g_2}{2}\hat{a}\ket{e}_j\bra{\uparrow} + \; \text{H.c.} \right).   
\label{eqn:ham_full_1}
\end{eqnarray}

The drive laser frequencies are arranged such that $\omega_1 = \omega_c + \omega_0$ and $\omega_2 = \omega_c - \omega_0$. We assume that the splitting $\omega_{\downarrow e} - \omega_{\uparrow e}$ between the spin states, nominally $\omega_0$, can be slightly modified, e.g. by a weak external magnetic field that we wish to sense, i.e. $\omega_{\downarrow e} - \omega_{\uparrow e} = \omega_0 + 2 \delta$, where $\delta$ ($-\delta$) is the shift of the $\ket{\uparrow}$ ($\ket{\downarrow}$) state.      

The detunings of the drive lasers from the atomic transitions are given by $\Delta_1 = \omega_1 - \omega_{\downarrow e} = -\Delta + \omega_0/2 - \delta$ and $\Delta_2 = \omega_2 - \omega_{\uparrow e} = -\Delta - \omega_0/2 + \delta$. Similarly, the detunings of the cavity mode from the atomic transitions are $\Delta_1^c = \omega_c - \omega_{\downarrow e} = -\Delta -\omega_0/2 - \delta$ and $\Delta_2^c = \omega_c - \omega_{\uparrow e} = -\Delta + \omega_0/2 + \delta$. 

We write the interaction Hamiltonian expressing the energy requirements in Eq.~(\ref{eqn:ham_full_1}) using complex exponentials involving these detunings as 
\begin{eqnarray}
&&\hat{H}_I(t) = \nonumber\\
&&\hphantom{+} \sum_j \left( \frac{\Omega_1}{2}\ket{e}_j\bra{\downarrow}e^{-i\Delta_1 t}                    
+\frac{\Omega_2}{2}\ket{e}_j\bra{\uparrow}e^{-i\Delta_2 t} + \; \text{H.c.}\right) \nonumber\\
&&+ \sum_j \left( \frac{g_1}{2}\hat{a}\ket{e}_j\bra{\downarrow}e^{-i\Delta_1^c t} 
+ \frac{g_2}{2}\hat{a}\ket{e}_j\bra{\uparrow}e^{-i\Delta_2^c t} + \; \text{H.c.} \right).\nonumber\\   
\end{eqnarray}

We use the effective Hamiltonian theory of Ref. \cite{james2007CJPDup} to derive the effective Hamiltonian in the limit where the detunings are all much greater than the Rabi frequencies. This effective Hamiltonian has three parts 
\begin{equation}
    \hat{H}_\text{eff}(t) = \hat{H}_\text{Stark}(t) + \hat{H}_\text{atom-atom}(t) + \hat{H}_\text{Raman}(t),  
\end{equation}

\noindent where 
\begin{eqnarray}
&&\hat{H}_\text{Stark}(t) = \sum_j \frac{\abs{\Omega_1}^2}{4\Delta_1}\left( \ket{\downarrow}_j\bra{\downarrow} - \ket{e}_j\bra{e} \right)  \nonumber\\
&& + \sum_j \frac{g_1^2}{4\Delta_1^c} \left( \hat{a}^\dag\hat{a} \left(\ket{\downarrow}_j\bra{\downarrow}-\ket{e}_j\bra{e} \right)  -\ket{e}_j\bra{e} \right) \nonumber\\
&& + \sum_j  \frac{\Omega_1 g_1}{4h(\Delta_1,\Delta_1^c)} \hat{a}^\dag \left( \ket{\downarrow}_j\bra{\downarrow}-\ket{e}_j\bra{e}\right)e^{i(\Delta_1^c-\Delta_1)t} + \; \text{H.c.} \nonumber \\
&& + \; \downarrow \rightarrow \uparrow \;(1\rightarrow 2),
\end{eqnarray}

\begin{eqnarray}
&&\hat{H}_\text{atom-atom}(t) = \nonumber\\
&& -\sum_{j,k\neq j} \frac{g_1^2}{4\Delta_1^c} \ket{e}_j\bra{\downarrow} \otimes \ket{\downarrow}_k\bra{e} + \; \downarrow \rightarrow \uparrow \;(1\rightarrow 2) \nonumber\\
&& -\sum_{j,k\neq j} \frac{g_1 g_2}{4h(\Delta_1^c,\Delta_2^c)}\left( \ket{\uparrow}_j\bra{e} \otimes \ket{e}_k\bra{\downarrow}e^{i(\Delta_2^c-\Delta_1^c)t} + \; \text{H.c.}  \right), \nonumber\\
\end{eqnarray}

\noindent and
\begin{eqnarray}
&&\hat{H}_\text{Raman}(t) = \sum_j  \frac{\Omega_1\Omega_2^*}{4h(\Delta_1,\Delta_2)}\ket{\uparrow}_j\bra{\downarrow}e^{i(\Delta_2-\Delta_1)t} + \; \text{H.c.}  \nonumber\\
&&\hphantom{\hat{H}_\text{Raman}(t)} + \sum_j \frac{g_1 g_2}{4h(\Delta_1^c,\Delta_2^c)} \hat{a}^\dag\hat{a} \ket{\uparrow}_j\bra{\downarrow}e^{i(\Delta_2^c-\Delta_1^c)t} + \; \text{H.c.} \nonumber\\
&&\hphantom{\hat{H}_\text{Raman}(t)} + \sum_j \frac{\Omega_1 g_2}{4h(\Delta_1,\Delta_2^c)}\hat{a}^\dag \ket{\uparrow}_j\bra{\downarrow}e^{i(\Delta_2^c-\Delta_1)t} + \; \text{H.c.} \nonumber\\
&&\hphantom{\hat{H}_\text{Raman}(t)} + \sum_j \frac{\Omega_2 g_1}{4h(\Delta_2,\Delta_1^c)}\hat{a}^\dag \ket{\downarrow}_j\bra{\uparrow}e^{i(\Delta_1^c-\Delta_2)t} + \; \text{H.c.} \nonumber\\
\end{eqnarray}

In the above expressions, $h(a,b) = 2/(a^{-1}+b^{-1})$ is the harmonic mean of $a$ and $b$. All terms in the effective Hamiltonian conserve the number of excitations in $\ket{e}$. This means that if the atoms are initially in the $\ket{\downarrow}-\ket{\uparrow}$ manifold, then the state $\ket{e}$ is negligibly populated and all interactions involving this level, and consequently, $\hat{H}_\text{atom-atom}(t)$,  can be dropped. Expressing the difference detunings in the complex exponentials in terms of $\Delta, \omega_0$, and $\delta$ shows the presence of rapidly oscillating terms with frequency $\sim \omega_0$ and slowly varying terms with zero frequency or a small frequency $\delta$. For $\Omega_1\approx \Omega_2 \sim \Omega$, the rapidly oscillating terms can be neglected since we operate in the regime where $\Omega^2/\Delta \ll \omega_0$. The resulting Hamiltonian consists of 
\begin{eqnarray}
&& \hat{H}_\text{Stark}(t) = \sum_j \frac{\abs{\Omega_1}^2}{4\Delta_1}\ket{\downarrow}_j\bra{\downarrow} + \sum_j \frac{g_1^2}{4\Delta_1^c}\hat{a}^\dag\hat{a}\ket{\downarrow}_j\bra{\downarrow} \nonumber\\
&& \hphantom{\hat{H}_\text{Stark}(t)} + \downarrow\rightarrow\uparrow \; (1\rightarrow 2)
\end{eqnarray}

\noindent and
\begin{eqnarray}
&&\hat{H}_\text{Raman}(t) = \sum_j \frac{\Omega_1 g_2}{4h(\Delta_1,\Delta_2^c)}\hat{a}^\dag \ket{\uparrow}_j\bra{\downarrow}e^{2i\delta t} + \; \text{H.c.} \nonumber\\
&&\hphantom{\hat{H}_\text{Raman}(t)} + \sum_j \frac{\Omega_2 g_1}{4h(\Delta_2,\Delta_1^c)}\hat{a}^\dag \ket{\downarrow}_j\bra{\uparrow}e^{-2i\delta t} + \; \text{H.c.} \nonumber\\
\end{eqnarray}

\subsection{Simple picture}

For $\Delta \gg \omega_0$, we can make the substitution $\Delta_1,\Delta_2,\Delta_1^c,\Delta_2^c\rightarrow -\Delta$. Then, with $\Omega_1 = \Omega_2 = \Omega_0$, the Stark shifts from the drive lasers shift the two spin states identically and therefore lead to an overall energy shift of $-N\Omega_0^2/4\Delta$, where we assume $\Omega_0$ is real. Similarly, with $g_1=g_2=g_0$, the frequency of the cavity mode is shifted by an amount $-Ng_0^2/4\Delta$ on account of the atom-cavity interaction. This can be compensated for by shifting the frequency of the drive lasers by the same amount. Introducing the collective angular momentum operators $\hat{J}_+ \equiv \sum_j\ket{\uparrow}_j\bra{\downarrow}$, $\hat{J}_- \equiv \hat{J}_+^\dag$, and $\hat{J}_z \equiv \sum_j \left(\ket{\uparrow}_j\bra{\uparrow}-\ket{\downarrow}_j\bra{\downarrow}\right)/2$, we can express the effective Hamiltonian as 
\begin{equation}
    \hat{H}_\text{eff} = 2\delta \hat{J}_z + \frac{\Omega_0 g_0}{4\Delta}\left(\hat{a}+\hat{a}^\dag\right)\left(\hat{J}_+ + \hat{J}_-\right), 
    \label{eqn:eff_ham_final}
\end{equation}

\noindent where we have let $\Omega_0 \rightarrow -\Omega_0$. The second term on the RHS is precisely the QND Hamiltonian described in Eq.~(1) of the Main Text. This coarse-grained Hamiltonian is valid over time intervals $\Delta t\gg T_{\omega_0} \equiv 2\pi/\omega_0$, and therefore, we require $\delta \ll \omega_0$ and that $\delta$ is approximately constant over the interval $\Delta t$. Mathematically, the latter implies $d\ln\delta/dt \ll 1/\Delta t \ll \omega_0/2\pi$.  

\subsection{Accounting for $\omega_0/\Delta$}

For $\delta \approx 0$, $h(\Delta_1,\Delta_2^c)=\Delta_1$ and $h(\Delta_2,\Delta_1^c)=\Delta_2$. To isolate the balanced cavity-assisted Raman transitions, three requirements have to be satisfied \cite{dimer2007PRADup}:

\begin{enumerate}[label={(\arabic*)}]
    \item Equal drive laser Stark shifts on both spin states: $\Omega_1^2/4\Delta_1=\Omega_2^2/4\Delta_2$. \label{r1}
    \item Equal frequency shift of cavity mode per atom in either spin state: $g_1^2/4\Delta_1^c=g_2^2/4\Delta_2^c$. \label{r2}
    \item Balanced Raman transitions: $\Omega_1 g_2/4\Delta_1 = \Omega_2 g_1/4\Delta_2$. \label{r3}
\end{enumerate}

We note that arranging $\Omega_1/\Omega_2$ and $g_1/g_2$ to satisfy \ref{r1} and \ref{r2} above automatically results in satisfying requirement \ref{r3}.

\subsection{Note concerning drive laser frequencies}

In practice, the frequency arrangement  of the drive lasers requires their average frequency $\omega_\text{av}$ to be tuned well within the cavity linewidth, i.e. $\abs{\omega_c-\omega_\text{av}}\ll\kappa$. The difference frequency $\omega_1-\omega_2$ is relatively easier to stabilize, and deviations from $2\omega_0$ manifest as a growth of the phase over time that can be measured and statistically modeled.

\section{\label{sec:fss} Phenomenological free-space scattering model}

\begin{figure}[!htb]
    \centering
    \includegraphics[width=\columnwidth]{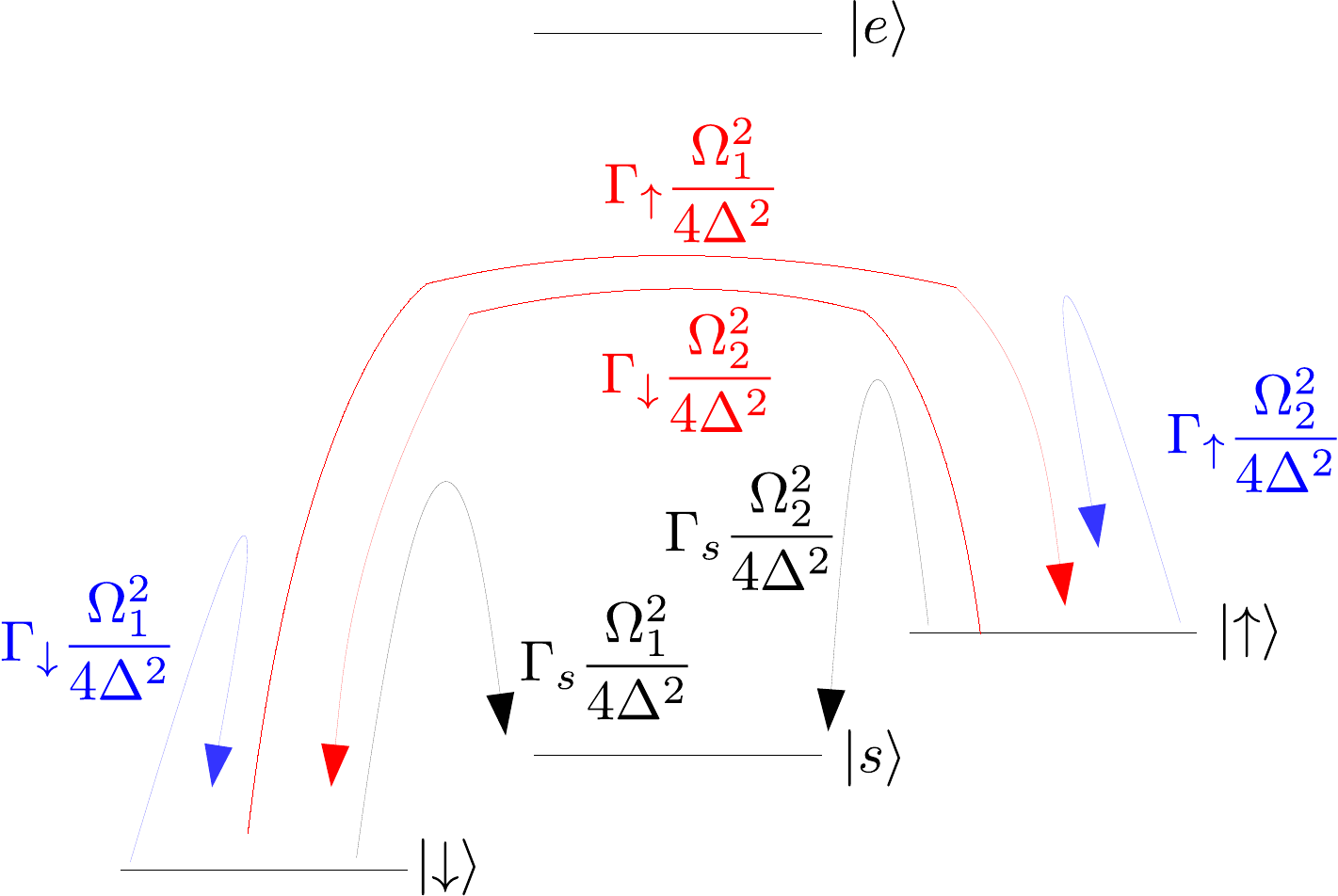}
    \caption{Free-space scattering processes considered in our model: Rayleigh scattering (blue), Raman scattering (red) and atom loss (black).}
    \label{fig:fss_schematic}
\end{figure}

Fig.~\ref{fig:fss_schematic} shows the various free-space scattering (FSS) processes considered in our model. The lifetime of the excited state is $\Gamma = \Gamma_{\downarrow}+\Gamma_{\uparrow}+ \Gamma_{s}$.

The dephasing rate is set by the total rate of Rayleigh scattering \cite{shankar2017PRADup}: 
\begin{equation}
    \text{Dephasing: \;} \Gamma_{\downarrow}\frac{\Omega_1^2}{4\Delta^2} + \Gamma_{\uparrow}\frac{\Omega_2^2}{4\Delta^2}. 
\end{equation}

Raman spin flips ($\ket{\downarrow}\rightarrow\ket{\uparrow}$) occur with rate $\Gamma_{\uparrow}\frac{\Omega_1^2}{4\Delta^2}$, while $\ket{\uparrow}\rightarrow\ket{\downarrow}$ occur at rate $\Gamma_{\downarrow}\frac{\Omega_2^2}{4\Delta^2}$ \cite{shankar2017PRADup}. Finally, atom loss occurs at a net rate given by 
\begin{equation}
    \text{Atom loss: \;} \Gamma_{s}\left(\frac{\Omega_1^2}{4\Delta^2}+\frac{\Omega_2^2}{4\Delta^2}\right).
\end{equation}

With $\Omega_1\approx \Omega_2 \equiv \Omega_0$, the probabilities for the different FSS channels are  
\begin{eqnarray}
r_d = \frac{\Gamma_{\downarrow}+\Gamma_{\uparrow}}{2\Gamma}, \;
r_{\downarrow\uparrow} = \frac{\Gamma_{\uparrow}}{2\Gamma}, \;
r_{\uparrow\downarrow} = \frac{\Gamma_{\downarrow}}{2\Gamma}, \;
r_l = \frac{\Gamma_{s}}{\Gamma}.
\label{eqn:probs}
\end{eqnarray}

The total FSS rate is 
\begin{equation}
    \gamma_\text{sc} = 2\Gamma \frac{\Omega_0^2}{4\Delta^2}.
     \label{eqn:gamma_sc} 
\end{equation}

The ``symmetric loss" model we adopt while presenting our numerical results arises in the case when $\Gamma_\downarrow=\Gamma_\uparrow=\Gamma_s = \Gamma/3$. In this case, from Eq.~(\ref{eqn:probs}), $r_d = 1/3$, $r_{\downarrow\uparrow}=r_{\uparrow\downarrow}=1/6$ and $r_l=1/3$.

Our simple phenomenological model captures the FSS processes expected when a three-level system is driven by two lasers whose difference frequency is far detuned from the two-photon resonance. In practice, scattering from any additional excited states should be considered. Detailed modeling of a real experiment will benefit from a rigorous first-principles derivation of the effects of free-space scattering that also accounts for the cavity mode, as shown in Refs.  \cite{borregaard2017NJPDup,sorensen2002PRADup}.

\vspace*{-1pc}

\subsection{Cooperativity}
In our theory, we have defined the cooperativity $C$ in terms of the two-photon rates $\Omega_\text{QND}$ and $\gamma_\text{sc}$. With the explicit form of $\gamma_\text{sc}$ in Eq.~(\ref{eqn:gamma_sc}), and the expression $\Omega_\text{QND} = \sqrt{2}\Omega_0 g_0/\Delta$, we can express $C$ in terms of the single-photon coupling strength $g_0$ and the lifetime $\Gamma$ of the state $\ket{e}$ as 
\begin{equation}
    C = 2\frac{\Omega_\text{QND}^2}{\kappa\gamma_\text{sc}} = 8\frac{g_0^2}{\kappa\Gamma}.
\end{equation}

We note that our definition of $C$ differs from the usual definition \cite{meiser2009PRLDup} by a factor of 8. However, as Eq.~(\ref{eqn:diff_eqn_cond_var}) shows, with our definition of $C$, the rate $\eta C\gamma_\text{sc}$ takes on the simple interpretation as the rate at which the variance in $J_x$ decreases as the field leaking out from the cavity is monitored. Here, $\eta$ is as usual the detection efficiency of the photodetectors used for homodyne detection. With perfect detectors ($\eta=1$), $C$ is therefore the ratio of the rates of the desirable and undesirable processes.   

\newpage

\section{\label{sec:gauss}Equations of motion of operator means and covariances}

Our starting point is Eq.~(3) of the Main Text, which we rewrite as 
\begin{eqnarray}
d \rho & = & \left(-i/\hbar[\hat{H}_\text{QND},\rho] + \kappa \mD[\hat{a}]\rho + \gamma_\text{sc} \sum_{j=1}^N \lv_1^j\rho \right) dt \nonumber\\
             &+& \sqrt{\eta \kappa} dW(t) \left(i\rho\hat{a}^\dag -i\hat{a}\rho -\sqrt{2}\ev{\hat{Y}}\rho \right),
             \label{eqn:me_diff}
\end{eqnarray}

\noindent where $dW(t)$ is a Wiener increment that satisfies $\overline{dW(t)}=0$ and $dW(t)^2 = dt$ \cite{thomsen2002JPhysBDup}. 

The expectation value of an operator $\hat{O}$ is given by $\ev{\hat{O}} = \text{Tr}\left[{\hat{O}\rho}\right]$ and, consequently, $\ev{\dot{\hat{O}}} = \text{Tr}\left[{\hat{O}\dot{\rho}}\right]$. We use a Gaussian approximation, i.e. keep track of only the means and covariances of the five operators $\hat{X}, \hat{Y},\hat{J}_x,\hat{J}_y$ and $\hat{J}_z$. We truncate these evolution equations at second order by factorizing third order moments of the type $\ev{\hat{O}_1 \hat{O}_2 \hat{O}_3}$ as 
\begin{eqnarray}
\ev{\hat{O}_1 \hat{O}_2 \hat{O}_3} &\approx& \ev{\hat{O}_1 \hat{O}_2} \ev{\hat{O}_3} + \ev{\hat{O}_2 \hat{O}_3}  \ev{\hat{O}_1} \nonumber\\ 
&+& \ev{\hat{O}_1 \hat{O}_3} \ev{\hat{O}_2} - 2 \ev{\hat{O}_1}\ev{\hat{O}_2}\ev{\hat{O}_3}.
\end{eqnarray}

This procedure leads to five equations governing the means of the operators that have the typical form of stochastic differential equations: $d\ev{\hat{O}} = a dt + b dW$. The time evolution of the fifteen covariances, on the other hand, are governed by ordinary differential equations that have no terms proportional to $dW$. This structure is a direct consequence of the Gaussian approximation we employ. We reproduce these equations below.

We use the notation $\ev{\ldots}$ to denote means and $\ev{\ldots}_c$ to denote (co)variances evaluated using the stochastic master equation, Eq.~(\ref{eqn:me_diff}). The covariance is defined as $\ev{\hat{O}_1\hat{O}_2}_c = (\ev{\hat{O}_1\hat{O}_2}+ \ev{\hat{O}_2\hat{O}_1})/2 - \ev{\hat{O}_1}\ev{\hat{O}_2}$.
%\vspace*{\fill}
%\clearpage
\vspace*{-2pc}

\begin{widetext}
\subsection{Operator means}

\begin{eqnarray}
d\ev{\hat{X}} &=& -\frac{\kappa}{2}\ev{\hat{X}} dt + 2 \sqrt{\frac{\eta \kappa}{2}}\ev{\hat{X}\hat{Y}}_c dW \nonumber\\
d\ev{\hat{Y}} &=& -\left(\frac{\kappa}{2} \ev{\hat{Y}} + \frac{\OQ}{2}\ev{\hat{J}_x}\right) dt + \sqrt{\frac{\eta \kappa}{2}} \left( 2\ev{\hat{Y}^2}_c - 1\right) dW \nonumber\\
d\ev{\hat{J}_x} &=& -\frac{\gamma_\text{sc}}{2} \ev{\hat{J}_x} dt +  2 \sqrt{\frac{\eta \kappa}{2}} \ev{\hat{Y}\hat{J}_x}_c dW \nonumber\\
d\ev{\hat{J}_y} &=& -\left[\frac{\gamma_\text{sc}}{2} \ev{\hat{J}_y} +\frac{\OQ}{2}\left(\ev{\hat{X}\hat{J}_z}_c + \ev{\hat{X}}\ev{\hat{J}_z}\right)\right]dt + 2 \sqrt{\frac{\eta \kappa}{2}} \ev{\hat{Y}\hat{J}_y}_c dW \nonumber\\
d\ev{\hat{J}_z} &=& -\left[\gamma_\text{sc} \left( \left(r_{\uparrow \downarrow} + r_{\downarrow \uparrow} + \frac{r_l}{2}\right)\ev{\hat{J}_z} + \left(r_{\uparrow \downarrow} -r_{\downarrow \uparrow}\right)\frac{N}{2} \right) - \frac{\OQ}{2}\left(\ev{\hat{X}\hat{J}_y}_c + \ev{\hat{X}}\ev{\hat{J}_y}\right) \right] dt + 2 \sqrt{\frac{\eta \kappa}{2}} \ev{\hat{Y}\hat{J}_z}_c dW \nonumber\\
\end{eqnarray}
\clearpage
\subsection{\label{subsec:covar}Operator covariances}

\begin{eqnarray}
\frac{d}{dt}\ev{\hat{X}^2}_c &=& -\kappa \left( \ev{\hat{X}^2}_c - \frac{1}{2}\right) - 2\eta \kappa \ev{\hat{X}\hat{Y}}_c^2 \nonumber\\
\frac{d}{dt}\ev{\hat{Y}^2}_c &=& -\kappa \left( \ev{\hat{Y}^2}_c - \frac{1}{2}\right) - \OQ \ev{\hat{Y}\hat{J}_x}_c - \frac{\eta \kappa}{2} \left(2\ev{\hat{Y}^2}_c-1\right)^2 \nonumber\\
\frac{d}{dt}\ev{\hat{X}\hat{Y}}_c &=& -\kappa \ev{\hat{X}\hat{Y}}_c - \frac{\OQ}{2}\ev{\hat{X}\hat{J}_x}_c - \eta \kappa \ev{\hat{X}\hat{Y}}_c \left(2\ev{\hat{Y}^2}_c-1\right)
\end{eqnarray}

\begin{eqnarray}
\frac{d}{dt}\ev{\hat{X}\hat{J}_x}_c &=& -\left(\frac{\kappa+\gamma_\text{sc}}{2}\right)\ev{\hat{X}\hat{J}_x}_c - 2 \eta \kappa \ev{\hat{Y}\hat{J}_x}_c \ev{\hat{X}\hat{Y}}_c \nonumber\\
\frac{d}{dt}\ev{\hat{X}\hat{J}_y}_c &=&
-\left(\frac{\kappa+\gamma_\text{sc}}{2}\right)\ev{\hat{X}\hat{J}_y}_c -\frac{\OQ}{2} \left(\ev{\hat{X}\hat{J}_z}_c\ev{\hat{X}} + \ev{\hat{X}^2}_c\ev{\hat{J}_z} \right) - 2\eta\kappa \ev{\hat{Y}\hat{J}_y}_c\ev{\hat{X}\hat{Y}}_c \nonumber\\
\frac{d}{dt}\ev{\hat{X}\hat{J}_z}_c &=& -\left(\frac{\kappa}{2} + \gamma_\text{sc}  \left(r_{\uparrow \downarrow} + r_{\downarrow \uparrow} + \frac{r_l}{2}\right)\right)\ev{\hat{X}\hat{J}_z}_c +\frac{\OQ}{2} \left(\ev{\hat{X}\hat{J}_y}_c\ev{\hat{X}} + \ev{\hat{X}^2}_c\ev{\hat{J}_y} \right) - 2\eta\kappa \ev{\hat{Y}\hat{J}_z}_c\ev{\hat{X}\hat{Y}}_c \nonumber\\
\end{eqnarray}

\begin{eqnarray}
\frac{d}{dt} \ev{\hat{Y}\hat{J}_x}_c &=& -\left(\frac{\kappa+\gamma_\text{sc}}{2}\right)\ev{\hat{Y}\hat{J}_x}_c 
-\frac{\OQ}{2}\ev{\hat{J}_x^2}_c -\eta\kappa \ev{\hat{Y}\hat{J}_x}_c \left( 2\ev{\hat{Y}^2}_c - 1\right) \nonumber\\
\frac{d}{dt} \ev{\hat{Y}\hat{J}_y}_c &=& -\left(\frac{\kappa+\gamma_\text{sc}}{2}\right)\ev{\hat{Y}\hat{J}_y}_c 
-\frac{\OQ}{2}\left(\ev{\hat{J}_x\hat{J}_y}_c + \ev{\hat{X}\hat{Y}}_c \ev{\hat{J}_z} + \ev{\hat{Y}\hat{J}_z}_c\ev{\hat{X}}\right) -\eta\kappa \ev{\hat{Y}\hat{J}_y}_c \left( 2\ev{\hat{Y}^2}_c - 1\right) \nonumber\\
\frac{d}{dt} \ev{\hat{Y}\hat{J}_z}_c &=& -\left(\frac{\kappa}{2} + \gamma_\text{sc}  \left(r_{\uparrow \downarrow} + r_{\downarrow \uparrow} + \frac{r_l}{2}\right)\right)\ev{\hat{Y}\hat{J}_z}_c 
-\frac{\OQ}{2}\left(\ev{\hat{J}_z\hat{J}_x}_c - \ev{\hat{X}\hat{Y}}_c \ev{\hat{J}_y} - \ev{\hat{Y}\hat{J}_y}_c\ev{\hat{X}}\right) -\eta\kappa \ev{\hat{Y}\hat{J}_z}_c \left( 2\ev{\hat{Y}^2}_c - 1\right) \nonumber\\
\end{eqnarray}

\begin{eqnarray}
\frac{d}{dt} \ev{\hat{J}_x^2}_c &=& -\gamma_\text{sc} \left( \ev{\hat{J}_x^2}_c - \left( r_{\uparrow \downarrow} + r_{\downarrow \uparrow} + r_d + \frac{r_l}{2}\right)\frac{N}{4} \right) -2 \eta\kappa \ev{\hat{Y}\hat{J}_x}_c^2 \nonumber\\
\frac{d}{dt} \ev{\hat{J}_y^2}_c &=& -\gamma_\text{sc} \left( \ev{\hat{J}_y^2}_c - \left( r_{\uparrow \downarrow} + r_{\downarrow \uparrow} + r_d + \frac{r_l}{2}\right)\frac{N}{4} \right) -\OQ \left(\ev{\hat{X}\hat{J}_y}_c \ev{\hat{J}_z} + \ev{\hat{J}_y\hat{J}_z}_c\ev{\hat{X}} \right) -2\eta\kappa \ev{\hat{Y}\hat{J}_y}_c^2 \nonumber\\
\frac{d}{dt} \ev{\hat{J}_z^2}_c &=& -\gamma_\text{sc} \left( \left(2r_{\uparrow \downarrow} + 2r_{\downarrow \uparrow} + r_l \right) \ev{\hat{J}_z^2}_c - r_{\uparrow \downarrow}\left(\frac{N}{2} + \ev{\hat{J}_z}\right) 
- r_{\downarrow \uparrow}\left(\frac{N}{2} - \ev{\hat{J}_z}\right) -r_l\frac{N}{8} \right) \nonumber\\
&+& \OQ \left(\ev{\hat{X}\hat{J}_z}_c \ev{\hat{J}_y} + \ev{\hat{J}_y\hat{J}_z}_c\ev{\hat{X}} \right) -2\eta\kappa \ev{\hat{Y}\hat{J}_z}_c^2
\end{eqnarray}

\begin{eqnarray}
\frac{d}{dt} \ev{\hat{J}_x\hat{J}_y}_c &=& -\gamma_\text{sc} \ev{\hat{J}_x\hat{J}_y}_c -\frac{\OQ}{2} \left(\ev{\hat{X}\hat{J}_x}_c \ev{\hat{J}_z} + \ev{\hat{J}_z \hat{J}_x}_c \ev{\hat{X}}\right) -2\eta\kappa \ev{\hat{Y}\hat{J}_x}_c \ev{\hat{Y}\hat{J}_y}_c \nonumber\\
\frac{d}{dt} \ev{\hat{J}_y\hat{J}_z}_c &=& -\gamma_\text{sc} \left( \left(\frac{3}{2}r_{\uparrow\downarrow} +\frac{3}{2}r_{\downarrow\uparrow}+\frac{r_d}{2}+r_l\right)\ev{\hat{J}_y\hat{J}_z}_c -\left(\frac{r_{\uparrow\downarrow}-r_{\downarrow\uparrow}}{2}\right) \ev{\hat{J}_y}\right) \nonumber\\
&+& \frac{\OQ}{2}\left(\ev{\hat{X}\hat{J}_y}_c \ev{\hat{J}_y} + \ev{\hat{J}_y^2}_c \ev{\hat{X}} - \ev{\hat{X}\hat{J}_z}_c \ev{\hat{J}_z} -\ev{\hat{J}_z^2}_c \ev{\hat{X}}\right) - 2\eta\kappa \ev{\hat{Y}\hat{J}_y}_c \ev{\hat{Y}\hat{J}_z}_c \nonumber\\
\frac{d}{dt} \ev{\hat{J}_z\hat{J}_x}_c &=& -\gamma_\text{sc} \left( \left(\frac{3}{2}r_{\uparrow\downarrow} +\frac{3}{2}r_{\downarrow\uparrow}+\frac{r_d}{2}+r_l\right)\ev{\hat{J}_z\hat{J}_x}_c -\left(\frac{r_{\uparrow\downarrow}-r_{\downarrow\uparrow}}{2}\right) \ev{\hat{J}_x}\right) \nonumber\\
&+&\frac{\OQ}{2} \left(\ev{\hat{X}\hat{J}_x}_c \ev{\hat{J}_y} + \ev{\hat{J}_x \hat{J}_y}_c \ev{\hat{X}}\right) -2\eta\kappa \ev{\hat{Y}\hat{J}_z}_c \ev{\hat{Y}\hat{J}_x}_c 
\end{eqnarray}

\end{widetext}

\subsection{Some comments on the equations of motion}

The above equations describe the conditional evolution when no external phase modulation is applied. Such a modulation can be straightforwardly accounted for by including the contributions of an additional Hamiltonian term $\propto \hat{J}_z$, such as the first term on the RHS of Eq.~(\ref{eqn:eff_ham_final}), to the equations of motion of the means and covariances.

In deriving the above equations,  we have accounted for the contribution of the atom loss terms (proportional to $r_l$) to the loss of coherence and the increase in diffusion. We note that atom loss reduces the effective number of atoms $N_\text{eff}$ that interact with the cavity mode. However, we simply use the total number $N$ wherever $N_\text{eff}$ explicitly appears in these equations. That is, we account for $\dot{N}_\text{eff} = -r_l \gamma_\text{sc} N_\text{eff}$ that leads to increased diffusion of the atomic spin components, but approximate $N_\text{eff} \approx N$ everywhere in the above equations.

\subsection{Numerical evolution}

We numerically evolve the dynamical equations for the means and covariances using a variation of the Improved Euler scheme for integrating stochastic differential equations \cite{roberts2012arxivDup}. Our numerical results are obtained using a C++ program that employs linear algebra features provided by Eigen \cite{eigenwebDup}, a C++ template library. The Wiener increments and random numbers required for numerical integration are obtained using random number generators from the GNU Scientific Library \cite{galassi2018scientificDup}.

\section{Analytic expression for variance in the difference measurement}

The time evolution of the conditional variance $\ev{\hat{J}_x^2}_c$ satisfies the Riccati equation 
\begin{equation}
    \frac{d}{dt}\ev{\hat{J}_x^2}_c = -\gamma_\text{sc}\left(\ev{\hat{J}_x^2}_c-\beta \frac{N}{4}\right)-\eta C\gamma_\text{sc} \ev{\hat{J}_x^2}_c^2,
    \label{eqn:diff_eqn_cond_var}
\end{equation}

\noindent where we have used the bad-cavity limit to adiabatically eliminate the coherence $\ev{\hat{Y}\hat{J}_x}_c \approx -(\Omega_\text{QND}/\kappa) \ev{\hat{J}_x^2}_c$ (see Section~\ref{subsec:covar}). In the regime where $\sqrt{NC}\gg 1$ and $\sqrt{\beta \eta NC} \gg 1$, the solution to Eq.~(\ref{eqn:diff_eqn_cond_var}) simplifies to 
\begin{equation}
    \ev{\hat{J}_x^2 (t)}_c = \frac{N}{4}\sqrt{\frac{4\beta}{NC\eta}}\coth
    \left[ \sqrt{\frac{\beta N C}{4\eta}}\gamma_\text{sc} t + \sqrt{\frac{4\beta}{NC\eta}} \right].
\end{equation}

The conditional mean $\ev{\hat{J}_x}$ is given by 
\begin{equation}
    \ev{\hat{J}_x (t)} = -\sqrt{\eta C \gamma_\text{sc}} \int_0^t dt' 
    \ev{\hat{J}_x^2 (t')}_c e^{-\frac{\gamma_\text{sc}}{2}(t-t')} \xi(t').
    \label{eqn:cond_jx}
\end{equation}

From Eq.~(5) and the bad-cavity relation $\ev{\hat{Y}}=-(\Omega_\text{QND}/\kappa)\ev{\hat{J}_x}$, the instantaneous photocurrent $i(t)$ carries information about $\ev{\hat{J}_x}$ but is corrupted by photon shot noise. Using Eq.~(6), the estimate $J_x^\text{(m)}$ from the photocurrent measured in an interval $[T_\text{i},T_\text{f}]$ is related to the conditional mean $\ev{\hat{J}_x}$ as 
\begin{equation}
    J_x^\text{(m)} = \frac{1}{T_\text{f} - T_\text{i}} \int_{T_\text{i}}^{T_\text{f}} dt 
    \left( \ev{\hat{J}_x (t)} - \frac{1}{\sqrt{\eta C \gamma_\text{sc}}} \xi(t) \right),
\end{equation}

%\vspace*{0.5pc}
\noindent where $\ev{\hat{J}_x}$ satisfies Eq.~(\ref{eqn:cond_jx}). In deriving Eq.~(8), we perform the first measurement $J_{x,1}^\text{(m)}(T_\text{W})$ over the window $[0,T_\text{W}]$, and the second measurement $J_{x,2}^\text{(m)}(T_\text{W})$ over the window $[T_\text{W},2T_\text{W}]$. The variance in the difference measurement $(\Delta J_{x,\text{diff.}}^\text{(m)})^2(T_\text{W})$ is given by 
\begin{equation}
    (\Delta J_{x,\text{diff.}}^\text{(m)})^2(T_\text{W}) = \overline{\left(J_{x,2}^\text{(m)}(T_\text{W})-J_{x,1}^\text{(m)}(T_\text{W})\right)^2},
\end{equation}

\noindent where the overbar indicates averaging over all possible realizations of the noise $\xi(t)$, which has the properties $\overline{\xi(t)} = 0$ and $\overline{\xi(t)\xi(t')}=\delta(t-t')$. We use the approximation that $\gamma_\text{sc} T_\text{W} \ll 1$ for the measurement windows we consider, so that the exponential factor in Eq.~(\ref{eqn:cond_jx}) can be set to unity. The resulting integrals can be evaluated analytically, resulting in 
\begin{equation}
    (\Delta J_{x,1}^\text{(m)})^2(T_\text{W}) = \frac{N}{4}\left( 1 + \frac{T_0}{T_\text{W}} + \frac{4\beta}{3\eta N C}\frac{T_\text{W}}{T_0} \right)
\end{equation}

\noindent for the variance in the first measurement, and 
\begin{equation}
    (\Delta J_{x,2}^\text{(m)})^2(T_\text{W}) = \frac{N}{4}\left( 1 + \frac{T_0}{T_\text{W}} + \frac{16\beta}{3\eta N C}\frac{T_\text{W}}{T_0} \right)
\end{equation}

\noindent for the variance in the second measurement, and the expression, Eq.~(8), for the variance in the difference measurement.

\subsection{Physical explanation for optimum measurement window}

Although the photon shot noise is averaged down as the measurement window $T_\text{W}$ is increased, undetected photons emitted via free-space scattering (FSS) lead to increased ignorance about the actual state of the collective spin. The photocurrent measurements in the initial parts of the measurement window are no longer as reliable in estimating the current value of $J_x$ as those in the latter parts, since FSS has significantly affected the collective spin state. Since  $J_{x,\text{diff.}}^\text{(m)}$ is the difference of measurements in two such windows, for very large $T_\text{W}$, the correlation in these two measurements decreases as a result of FSS. The upshot: The measurement window has an optimum $T_\text{opt}$ below which the measurement suffers from photon shot-noise, and above which it is affected by FSS.

\section{Atom number fluctuations}

\begin{figure}[!htb]
    \centering
    \includegraphics[width=\linewidth]{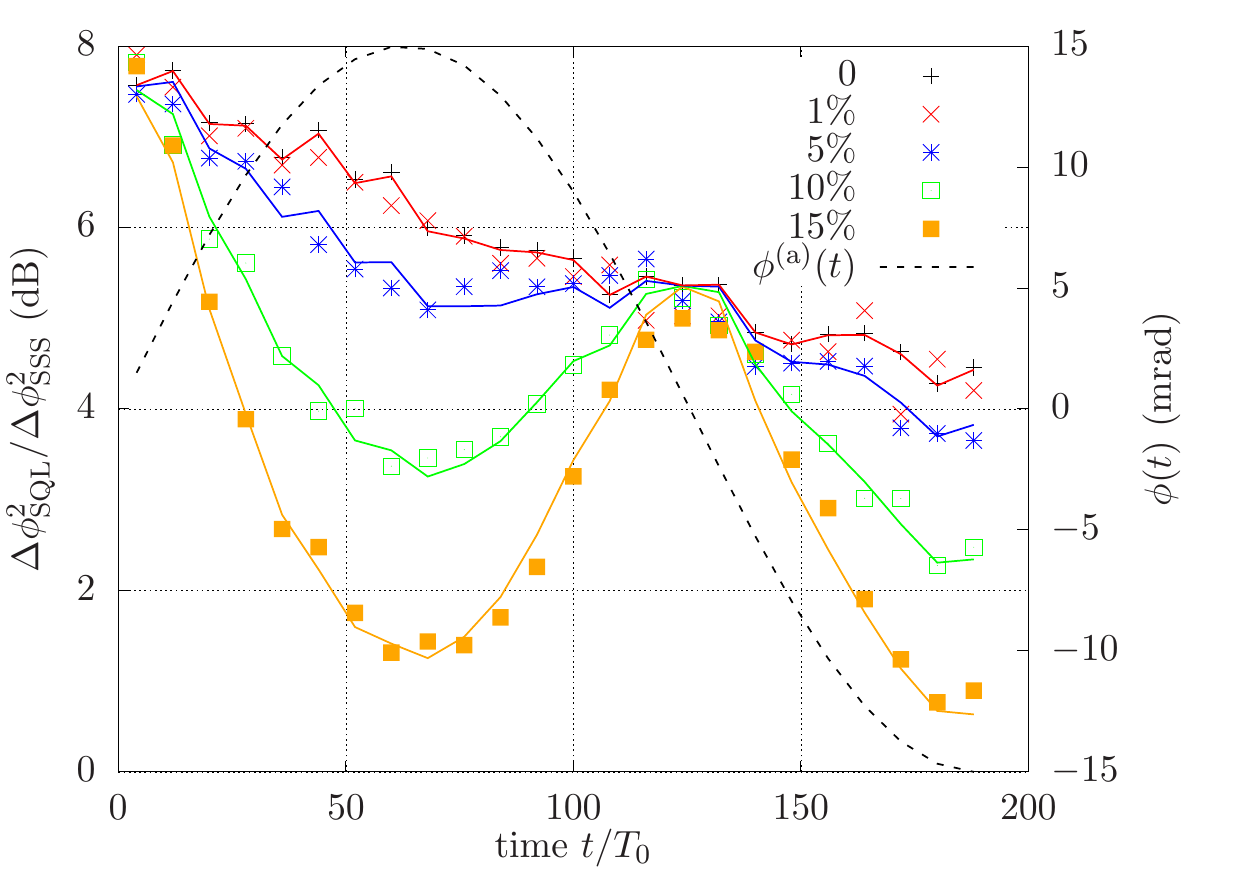}
    \caption{Effect of number fluctuations on real-time phase tracking. Single-run precision gain computed using error histograms of 2048 experimental runs (compare to Fig.~2(c) of the Letter). Parameters are from Fig.~2 of the Letter, except that the number of atoms in each run is variable, with a mean of $\bar{N}=10^5$ and $\Delta N/\bar{N}$ indicated by the percentages. Data points (markers) extracted from the numerical experiments are in very good agreement with semi-analytic results (solid lines) obtained using the simple expression in Eq.~(\ref{eqn:phi_error_2}) that accounts for number fluctuations. The black dashed line plots the applied phase modulation, for reference.}
    \label{fig:num_fluct}
\end{figure}

Here, we show why the number fluctuations are not important in a practical realization of our proposal. 

In the small angle limit, the measured phase is linearly related to $J_x^\text{(m)}$ as 
\begin{equation}
    \phi^\text{(m)} = \frac{2 J_x^\text{(m)}}{\vis N},
\end{equation}

\noindent where $\vis$ is the visibility and $N$ is the number of atoms. In the 
Main Text, we have assumed fluctuations only in $J_x^\text{(m)}$. In addition, when $N$ is 
not precisely known, the error in $\phi^\text{(m)}$ is given by 
\begin{equation}
    (\Delta \phi^\text{(m)})^2 \approx \left(\frac{2}{\vis N}\right)^2 (\Delta J_x^\text{(m)})^2 + (\phi^\text{(m)})^2 \left(\frac{\Delta N}{N}\right)^2.
    \label{eqn:phi_error_2}
\end{equation}

The expression in Eq.~(\ref{eqn:phi_error_2}) is approximate because we neglect the covariance of fluctuations in $J_x$ and $N$. The error from $\Delta N$ is negligible compared to the error from $\Delta J_x$ when 
\begin{equation}
    (\phi^\text{(m)})^2 \ll \frac{4}{\vis^2}\left(\frac{\Delta J_x^\text{(m)}}{\Delta N}\right)^2.
\end{equation}

In other words, the presence of number fluctuations only sets an upper bound on the dynamical range of the phase that can be tracked, and does not impose a fundamental restriction. 

In order to validate this argument, we introduce number fluctuations in our numerical experiments for the situation depicted in Fig. 2 of the Letter, see Fig.~\ref{fig:num_fluct}. When $\Delta N = 0$, we recover the results of Fig.~2(c) of the Main Text. As we increase $\Delta N$, the precision gain sharply falls in regions where the tracked phase has large amplitude. To verify the simple formula, Eq.~(\ref{eqn:phi_error_2}), we take the numerically obtained variance in the absence of number fluctuations (black markers) as representing the $(\Delta J_x^\text{(m)})^2$ term in Eq.~(\ref{eqn:phi_error_2}) and add, by hand, the contribution of the $(\Delta N)^2$ term, to obtain the solid lines shown in Fig.~\ref{fig:num_fluct}. The values of $\phi^{(m)}$ that enter this expression are approximated to be the values of the applied phase modulation (black dashed line) at the window centers. The very good agreement between the numerically extracted (markers) and semi-analytic (solid lines) results validate the expression, Eq.~(\ref{eqn:phi_error_2}), for the error in $\phi$.

Figure~\ref{fig:num_fluct} and Eq.~(\ref{eqn:phi_error_2}) imply that, in a practical realization, the effect of number fluctuations can be suppressed if the measured phase is always maintained close to zero. This goal can be achieved using the feedback scheme we mention in the conclusion of the Main Text. It enables continuous tracking of large phase excursions in a small angle measurement limit, and therefore greatly suppresses sensitivity to atom number fluctuations. In fact, by the same argument, a feedback loop will also suppress variations or uncertainties in any scale factors relating the measured current to the phase $\phi(t)$, and not just the atom number.

\bibliographystyle{apsrev4-1}

\begin{thebibliography}{53}
\bibitem [{vis()}]{visibility}%
  \BibitemOpen
  \bibinfo {note} {The Bloch vector is shortened by a factor $\mathcal{V}(t)
  \approx e^{-\gamma_\text{sc}t/2} e^{-(C\gamma_\text{sc}/4)t/2}$. For $C\ll
  1$, free-space scattering dominates, whereas for $C \gg 1$,
  measurement-induced squeezing is the dominant contribution.}\BibitemShut
  {Stop}%
\bibitem [{dcE()}]{dcErrorNote}%
  \BibitemOpen
  \bibinfo {note} {If only the form, and not the precise value, of an AC signal
  is to be tracked in a single run, the choice of initial phase reference is
  irrelevant, making the state preparation step unnecessary.}\BibitemShut
  {Stop}%
\bibitem [{jum()}]{jumpTimeNote}%
  \BibitemOpen
  \bibinfo {note} {For simplicity, we assume that the jump time of our test
  signal is known in advance. In practice, the jump time and amplitude can be
  simultaneously estimated using tools from optimal filtering
  theory.}\BibitemShut {Stop}%
\end{thebibliography}

\end{document}